\documentclass{webofc}
\usepackage[varg]{txfonts}   
\usepackage[table]{xcolor}
\usepackage{color}
\usepackage{colortbl}
\newcommand{\code}[1]{\textsc{#1}}
\def\TRENTo{{\sc t\kern-.05em \lower.5ex\hbox{r}\kern-.025em e\kern-.05em n\kern-.05em t\kern-.09em}o}
\graphicspath{ {figures/} }
\newcommand{\iccing}{\code{iccing}}
\newcommand{\ccake}{\code{ccake}}
\newcommand{\vUSPhydro}{\code{v-USPhydro}}

\begin{document}
\title{Influence of baryon number, strangeness, and electric charge fluctuations at the LHC}
%
%

\author{
\firstname{Christopher} \lastname{Plumberg}\inst{1}        \and
\firstname{Dekrayat} \lastname{Almaalol} \inst{2}\fnsep\thanks{\email{almaalol@illinois.edu}}                     \and
\firstname{Travis} \lastname{Dore} \inst{3}                \and
\firstname{Jordi} \lastname{Salinas San Mart\'in} \inst{2} \and
\firstname{Patrick} \lastname{Carzon} \inst{4}             \and
\firstname{Debora} \lastname{Mroczek} \inst{2}           \and
\firstname{Nanxi} \lastname{Yao} \inst{2}                  \and
\firstname{Willian} \lastname{M.~Serenone} \inst{2}           \and
\firstname{Lydia} \lastname{Spychalla} \inst{5}            \and
\firstname{Matthew} \lastname{Sievert} \inst{6}            \and
\firstname{Jacquelyn} \lastname{Noronha-Hostler} \inst{2}
}

\institute{
           Natural Science Division, Pepperdine University, Malibu, CA 90263, USA 
\and
Illinois Center for Advanced Studies of the Universe, Department of Physics, University of Illinois at Urbana-Champaign, Urbana, IL 61801, USA
\and
    Fakult\"at f\"ur Physik, Universit\"at Bielefeld,
D-33615 Bielefeld, Germany
\and 
            Department of Physics, Franciscan University, Steubenville, OH 43952, USA
\and 
            Pennsylvania State University,
Department of Meteorology and Atmospheric Science, State College, PA 16801, USA
\and
           Department of Physics, New Mexico State University, Las Cruces, NM 88003, USA
}

\abstract{%
  At the Large Hadron Collider it is possible to generate BSQ (baryon, strangeness, and electric) charge density fluctuations from gluon splittings into quark/anti-quark pairs, generated within the \iccing{} model. 
  In this work, we implement BSQ charge dynamics in a fully integrated framework. We propagate these conserved charges within an upgraded version of the \vUSPhydro{} hydrodynamic model, which conserves the BSQ densities exactly. Our hydrodynamic simulation uses the full 4D equation of state $\left\{T,\mu_B,\mu_S,\mu_Q\right\}$ from lattice Quantum Chromodynamics and includes decays from the Particle Data Group 2016+. We study the dynamical trajectories of fluid cells passing through the QCD phase diagram. We discuss future applications for this new framework.
}
\maketitle
\section{Introduction}
\label{sec:intro}

The Quark-Gluon Plasma (QGP) produced at top Large Hadron Collider (LHC) energies has negligible net-baryon number (B), strangeness (S), and electric charge (Q) at mid-rapidity.  However, due to gluons splitting into quark/anti-quark ($q\bar{q}$) pairs, one can get large local fluctuations of BSQ densities, which may affect the geometry of the initial state \cite{Carzon:2019qja,Carzon:2023zfp}. In this work, we take into account these local charge fluctuations using the \TRENTo+\iccing+\ccake{} framework that describes initial conditions, gluon splittings, and relativistic viscous hydrodynamics with BSQ conserved charges. We make use of the full 4-dimensional (4D) lattice Quantum Chromodynamics (QCD) equation of state (EoS) across finite temperatures, and baryon chemical potentials $\mu_B$, strangeness chemical potential $\mu_S$, and electrical charge chemical potential $\mu_Q$.  Here, we consider the role of the initial gluon splittings on both the dynamical evolution of the QGP and final state observables.  We find that early times reach large fluctuations in the chemical potentials which can remain significant at freeze-out.

\section{BSQ Hydrodynamics}
\label{sec:BSQhydro}

Our initial conditions are generated using \TRENTo{} \cite{Moreland:2014oya} (assuming the central values from \cite{Bernhard:2016tnd}). For each event, this gives an initial entropy density profile which we convert into energy density using the lattice QCD EoS \cite{Noronha-Hostler:2019ayj} and interpret as a distribution of pure gluons.  We then use \iccing{} \cite{Carzon:2019qja} to sample gluons from this initial distribution and split them into up, down, and strange $q\bar{q}$ pairs. This yields initial energy density $\varepsilon$, baryon density $\rho_B$, strangeness density $\rho_S$, and electric charge density $\rho_Q$ profiles which fluctuate from one event to the next.

For each event, the initial conditions are then fed into the relativistic viscous hydrodynamics code \ccake{} (Conserved ChArges HydrodynamiK Evolution) \cite{CCAKE}, which is based on an earlier code \vUSPhydro{} \cite{Noronha-Hostler:2013gga,Noronha-Hostler:2014dqa} that did not include conserved charge densities.
\ccake{} uses a Lagrangian algorithm known as Smooth Particle Hydrodynamics (SPH) to solve the Israel-Stewart equations with three conserved charges (currently neglecting diffusion; cf. \cite{Almaalol:2022pjc}).  The algorithm separates the initial conditions into fictitious ``SPH particles," each of which can be viewed as a locally equilibrated fluid cell in the simulation. 
We use the hydrodynamic parameters described in \cite{Alba:2017hhe} for the lattice QCD 2+1 EoS.
We assume a vanishing bulk viscosity $\zeta/s=0$ and constant shear viscosity $\eta/s=0.047$, implying a non-trivial $\eta T/(\varepsilon+p)$.
We have checked that \ccake{} reproduces well the known Gubser solutions \cite{Gubser:2010ze} for a finite shear viscosity \cite{Marrochio:2013wla} and for an ideal fluid with a conserved charge \cite{Gubser:2010ui,Denicol:2018wdp}.

The natural hydrodynamic variables in \ccake{} are entropy density $s$ and BSQ densities $\left\{\rho_B,\rho_S,\rho_Q\right\}$ but the  lattice QCD EoS is calculated on a tabulated 4D grid of $\left\{T,\mu_B,\mu_S,\mu_Q\right\}$ \cite{Noronha-Hostler:2019ayj}, and so one must employ interpolation and root-finding to use the EoS in \ccake{}.
In addition, the lattice QCD EoS is a Taylor series expansion in $\mu_X/T$ ($X=B,S,Q$) which begins to break down for values of $\mu_X/T\gtrsim 2.5$.
We therefore introduce a series of ``back-up" EoSs which mimic the lattice QCD EoS as closely as possible, either when it becomes unphysical (acausal or unstable), or when the root-finder yields degenerate solutions or no solution at all.
In this work, we include three back-up EoSs: (1) a hyperbolic tangent function that matches the lattice QCD table to a conformal EoS; (2) a conformal EoS; and (3), a ``diagonalized" conformal EoS which has  no cross terms between the temperature or chemical potentials.  Explicit forms for the back-up EoSs are provided in \cite{CCAKEpaper}.

At each time step in the hydrodynamic evolution, the $\left\{s,\rho_B,\rho_S,\rho_Q\right\}$ in each fluid cell are used to obtain a unique set of coordinates in $\left\{T,\mu_B,\mu_S,\mu_Q\right\}$.  For the occasional SPH particle (total $\lesssim 5\%$ in central collisions) for which this process fails, we instead obtain thermodynamics using one of the ``back-up" EoSs.  This permits a stable evolution while ensuring that the subsequent back-up EoSs play a negligible role in the overall dynamics. 

\begin{figure*}
    \centering
    \includegraphics[keepaspectratio, width=0.92\linewidth]{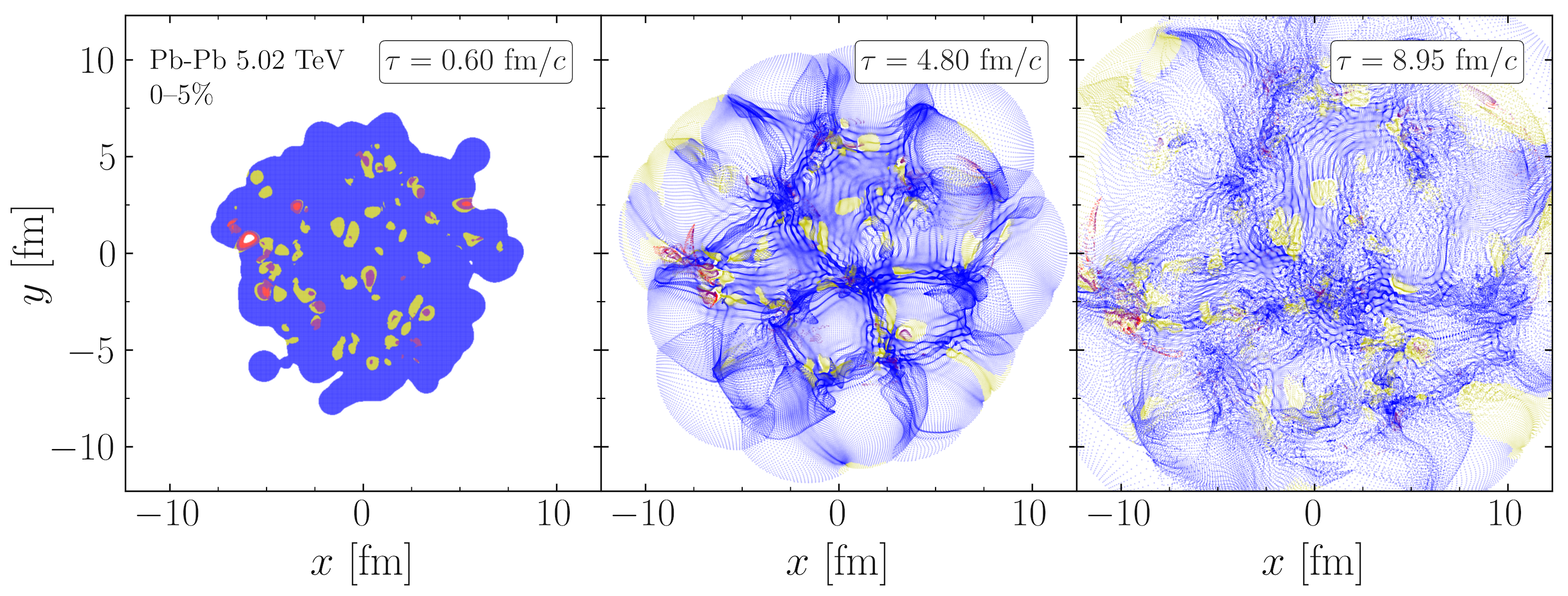}
    \caption{
    Equation of state (EoS) distribution of SPH particles on the transverse plane at times $\tau=0.6$, $5.25$, and $11.10$ fm$/c$ in a single event at the LHC for central collisions. SPH particles are color coded according to the EoS used: blue $=$ lattice QCD table, yellow $=$ hyperbolic $\tanh$ matching the conformal EoS to the lattice QCD table, purple $=$ conformal EoS, and red $=$ conformal-diagonal EoS.
    }
    \label{fig:EOSexpand}
\end{figure*}

In Fig.\ \ref{fig:EOSexpand} we show a central LHC event where individual fluid cells are color coded by EoS type at three different time steps.  The vast majority of the SPH particles use the lattice QCD table (shown in blue).  The back-up EoSs (1), (2), and (3) are ordered by decreasing frequency and are colored yellow, purple, and red (respectively).  We find that, even at late times, the vast majority of the SPH particles remain using the lattice QCD EoS.
One also observes in Fig.\ \ref{fig:EOSexpand} the behavior of individual SPH particles over time. The initial condition begins on a fixed grid, where a minimum energy density cutoff has been imposed.  The SPH particles follow trajectories based upon the equations of motion which lead to the formation of both high and low density regions at later times.  Note that the density of SPH particles need not correlate with the physical energy and charge densities in the system.

Using standard approaches, we implement freeze-out along a hypersurface of constant energy density $\varepsilon_{FO} = 266$ MeV/fm$^3$ that corresponds to a freeze-out temperature of $T_{FO}=150$ MeV at vanishing densities.  In this work we use the PDG2016+ \cite{Alba:2017mqu} particle list.  We have tuned the overall normalization constant of the initial conditions to reproduce all charge particle multiplicities at LHC Run 2 Pb+Pb collisions using ALICE data \cite{ALICE:2019hno}.

\section{Results: Exploration of the dense matter EoS at LHC energies }
\label{sec:results}

\begin{figure*}
    \centering
    \includegraphics[keepaspectratio, width=0.9\linewidth]{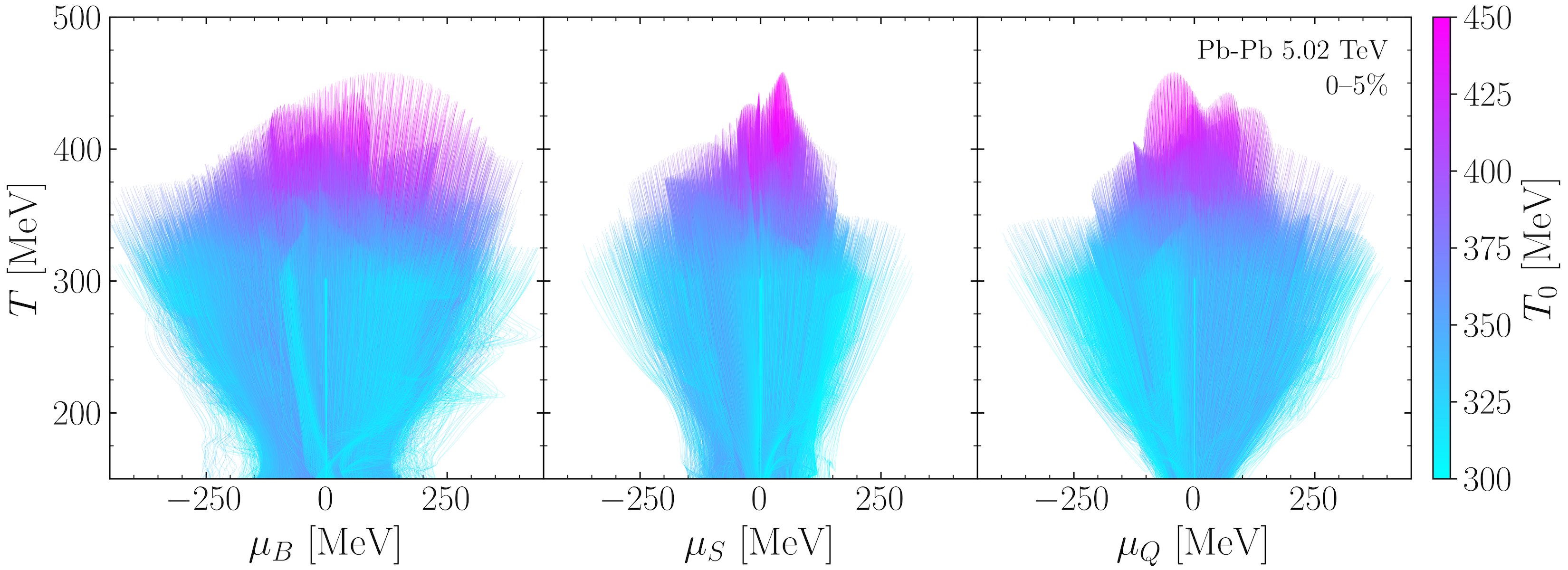}
    \caption{Trajectories of fluid cells that have initial temperatures above $T_0>300$ MeV across the phase diagram in $\left\{ T, \mu_B\right\}$ (left), $\left\{ T, \mu_S\right\}$ (middle), $\left\{ T, \mu_Q\right\}$ (right). The trajectories are color coded by their $T_0$. The BSQ local charge fluctuations survive during the hydrodynamical response and result in finite BSQ chemical potentials at the freeze-out hypersurface.
    }
    \label{fig:Tmu_trajs}
\end{figure*}

To explore the influence of local fluctuations in the charge densities, we plot the individual trajectories of various SPH particles across the QCD phase diagram.  Because we use SPH, it is possible to determine how a single fluid cell would flow over time across the QCD phase diagram in the space of $\left\{T,\mu_B,\mu_S,\mu_Q\right\}$.  Thus, we plot these trajectories in Fig.\ \ref{fig:Tmu_trajs} for SPH particles that have initial temperature $T_0$ above $T_0>350$ MeV. The choice in $T_0$ is arbitrary and is fixed here simply to make the system's behavior over time more easily discernible.  

Fig.\ \ref{fig:Tmu_trajs} reveals that the SPH particles begin at high $T$ in the phase diagram and subsequently evolve toward smaller $T$ and chemical potentials.  All panels show the particle trajectories as functions of temperature $T$ along the $y$-axis and $\mu_B$ (left), $\mu_S$ (center), and $\mu_Q$ (right) plotted along the $x$-axis. The lines are color coded by the initial temperature $T_0$ to help guide the eye.  In general, we find that the fluctuations in chemical potentials are largest at high temperatures, spanning nearly the entire available range in chemical potential.  Albeit the chemical potentials decrease with time, they remain significant even at freeze-out.

\section{Conclusions and Outlook}
\label{sec:concl}

We have developed a new framework to study BSQ conserved charges using the lattice QCD EoS and relativistic viscous hydrodynamics.  We study the passage of the system through the QCD phase diagram when including fluctuations of BSQ charges that arise from gluons splitting into $q\bar{q}$ pairs.  We observe broad fluctuations in charge densities even in regimes where they are expected to be zero on average.  We have also addressed the technical challenge of implementing a 4D EoS within the hydrodynamic framework by introducing ``back-up" EoSs to approximate the thermodynamics when the primary EoS becomes unusable.  Results for the particle multiplicities and collective flow will become public soon \cite{CCAKEpaper}. The results shown here open the door for future explorations of BSQ diffusion \cite{Greif:2017byw}, transport coefficients that depend on chemical potentials \cite{McLaughlin:2021dph}, and EoS studies at large densities. 

This work was supported by the US-DOE Nuclear Science Grant No. DESC0023861 and DE-SC0024560, the NSF Framework Grant No. OAC-2103680 for the MUSES collaboration, the Deutsche
Forschungsgemeinschaft under grant CRC-TR 211
“Strong-interaction matter under extreme conditions”
project no. 315477589-TRR 211, and  the Illinois Campus
Cluster, operated by the
Illinois Campus Cluster Program in conjunction
with the National Center for Supercomputing Applications that is supported by 
University of Illinois at Urbana-Champaign funds. 

\bibliography{inspire,NOTinspire}

%

\end{document}